\documentclass[aps, prl,twocolumn,superscriptaddress,showpacs,preprintnumbers,amsmath,amssymb]{revtex4-1}
\usepackage{amsmath}
\usepackage{amssymb}
\usepackage{dcolumn}% Align table columns on decimal point
\usepackage{graphicx}
\usepackage{bm}% bold math
\usepackage{epstopdf}
\usepackage{threeparttable}
\usepackage{float}
\usepackage[export]{adjustbox}

\begin{document}
%\begin{CJK}{UTF8}{gbsn} % Use default fonts from CJK (see below)
\title{Size-Sieving Separation of  Hard-Sphere Mixtures through Cylindrical Pores}

 \author{Yue Yu}
 \affiliation{Division of Natural and Applied Sciences, Duke Kunshan University, Kunshan, Jiangsu, 215300, China}

 \author{Kai Zhang  } % (\url{@.com})
\email{kai.zhang@dukekunshan.edu.cn}
 \affiliation{Division of Natural and Applied Sciences, Duke Kunshan University, Kunshan, Jiangsu, 215300, China}
  \affiliation{Data Science Research Center (DSRC), Duke Kunshan University, Kunshan, Jiangsu, 215300, China}

\date{\today}

\begin{abstract}
The collision dynamics of hard spheres and cylindrical pores is solved exactly, which is the minimal model for a regularly porous membrane.
 Nonequilibrium event-driven molecular dynamics simulations are used to show that the permeability  $P$ of hard spheres of size $\sigma$ through cylinderical pores of size $d$ follow the hindered diffusion mechanism  due to size exclusion as $P \propto (1-\sigma/d)^2$. Under this law,  the separation of binary mixtures of large and small particles  exhibits a linear relationship between $\alpha^{-1/2}$ and  $P^{-1/2}$, where $\alpha$ and $P$ are the selectivity  and permeability of the smaller particle, respectively. The mean permeability through polydisperse pores is the sum of permeabilities of individual pores, weighted by the fraction of the single pore area over the total pore area.  
\end{abstract}


\maketitle

Membrane separation is an energy-efficient  way to extract some substances from others  with a multitude of industrial applications such as water desalination, ion exchange, carbon capture and protein purification~\cite{baker2012}.  
The  goal  here  is to maximize the acquisition rate  of the desired species at the highest possible purity  by effectively filtering out undesired ones.  However, an intrinsic   trade-off   always exists  between this pair of separation performance characteristics -- permeability and selectivity~\cite{park2017}.  
%Depending on the problem, the particles to be separated can be angstrom-sized (\AA) ions/molecules, nanometer-sized ($n$m) macromolecules, micrometer-sized ($\mu$m) bacteria, etc. %The corresponding membranes used in these problems are often embedded with 
%The sizes of the pores or cavities in membranes need to be of similar length scales of the particles in order to efficiently separate them.
Although different penetrants, being  angstrom-sized (\AA) ions/molecules, nanometer-sized ($n$m) macromolecules, or micrometer-sized ($\mu$m) bacteria, may have different transport mechanisms,  it is now generally believed that  narrowing down the  size distribution of membrane pores can achieve higher separation performance~\cite{moon2020can,culp2021}. 
Apart from widely-used microfiltration and ultrafiltration membranes~\cite{mehta2005permeability},  tremendous efforts have been made to prepare regularly porous stuctures on nano- or molecular scales in systems such as  zeolites~\cite{smit2008},  metal organic frameworks~\cite{qiu2014} (MOFs), and silicon nanochannels~\cite{gruener2008}.  
Recent advances in self-assembly techniques  further enabled the synthesis of ceramic~\cite{kresge1992}, graphene-based~\cite{li2013} or block copolymer~\cite{phillip2006} molecular sieves,  which are featured with well-controlled  pores as regular as parallel cylinders. It is therefore needed to develop a quantitative understanding of how the steric (size) exclusion effect imposed by regular pores determines the separation of particles of different sizes.

In  the early study of  gas  transport  through long cylindrical tubes~\cite{knudsen1909}, it was found that the bulk self diffusivity $D_s = \frac{1}{3}\lambda \bar{v}$ of a gas with molecular mass $m$, mean free path $\lambda$  and mean velocity $\bar{v}  = \sqrt{\frac{ 8 k_B T}{  \pi m}}$ at temperature $T$  becomes  $D_K =  \frac{1}{3} d \bar{v}$ inside the tube, when the tube diameter $d \ll \lambda$. An implicit assumption of this Knudsen diffusion mechanism is that the diameter of  gas molecules $\sigma$ is much smaller than $d$ so that they can be treated as point masses. If two gases were to be separated under Knudsen diffusion, it is the mass dependence of $D_K \sim \frac{1}{\sqrt{m} }$ that makes the heavier one exit the tube later, regardless of their sizes. 
More theoretical analysis  showed that the transport diffusivity $D$ of  rarefied gases flowing under a density gradient   also depends on the length of the tube $L$ through the ratio $L/d$ and   $D$ approaches  $D_K$ only when $L \gg d$~\cite{pollard1948,shi2012,colson2019}. But their conclusions are again limited to the case  of $\sigma \to 0$.
In the context of molecular sieves, we are more interested in the regime of $\sigma  \lesssim d$ where size exclusion takes effect. Although modifications to the Knudsen diffusion theory have been proposed to replace $d$ with an effective diameter $d_{\rm eff} = d - \sigma$   expecting  that  $D_K \propto  d - \sigma$~\cite{gilron2002}, it  is not clear whether this rule actually holds and whether the transport  still  remains Knudsen-like.
 \begin{figure}[!b]%[H]%
\includegraphics[width=0.9\columnwidth]{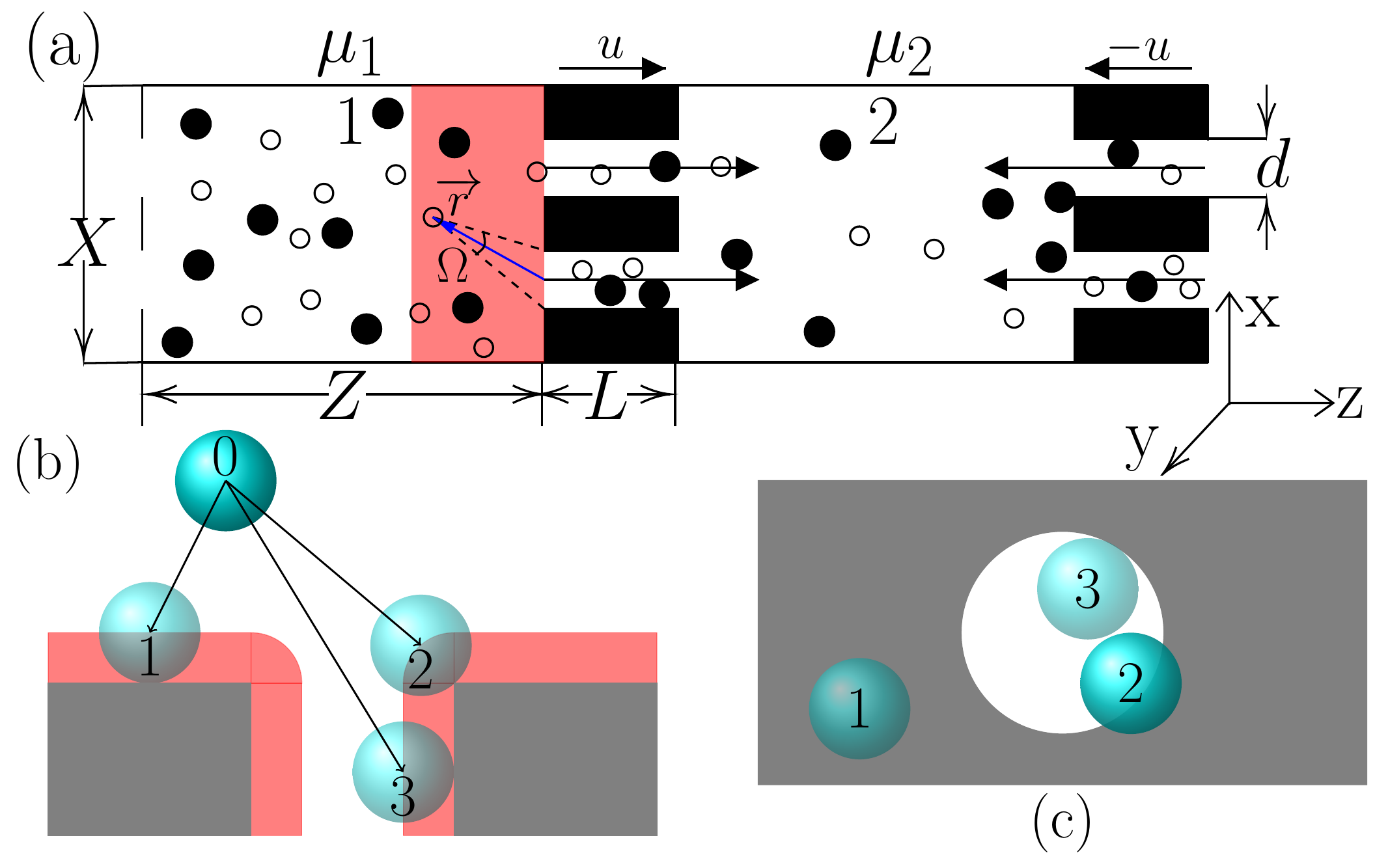}
\caption{(a) Simulation box setup of DCV-GCMD.  A particle at position ${\bf r}$ in the waiting zone (shaded)  makes a solid angle $\Omega$ to  the pore opening. (b) Side and (c) top view of the three possible cases of particle-pore collision events. }
\label{fig:method}
\end{figure}

In this Letter, we study the transport and separation of  low-density  binary hard-sphere mixtures through cylindrical pores under a chemical potential gradient (Fig.~\ref{fig:method}). Each mixture consists of  hard spheres $A$ (large) and $B$ (small)   flowing from the upstream chamber $1$ at chemical potential $\mu_1$ to the downstream chamber $2$ at a lower chemical potential $\mu_2$, implemented by the dual control volume grand canonical molecular dynamics (DCV-GCMD) method~\cite{heffelfinger1998massively,ford1998massively}. 
Previous  molecular simulations   on similar problems have  mostly focused on  Lennard-Jones like particles  under slit-shaped confinement~\cite{cracknell1995direct,furukawa1996non, furukawa1997, furukawa1997effects,xu1998nonequilibrium,xu2000nonequilibrium,takaba1998}, and in consequence, were not able to distinguish  size sieving from other effects. Other methods resort to equilibrium and/or stochastic sampling~\cite{mon2002,malek2003} that cannot genuinely reflect the nonequilibrium transport dynamics  driven by  pressure gradients.  To understand size-sieving filtration through  regularly porous membranes, it is  
imperative to faithfully simulate the entrance of particles into pores. We analytically resolve the collision dynamics of hard spheres and cylindrical pores by considering three possibilities (Fig.~\ref{fig:method}(b-c)). A hard sphere can hit the membrane surface and bounce back (case 1), or hit the circular edge of the pore and then bounce away (case 2),  or directly fly into the pore and hit its interior wall (case 3). We can compute the particle-pore collision time by considering the excluded volume of a sphere over the pore surface (red shaded area in Fig.~\ref{fig:method}(b)). The final solution is about finding the intersection between a linear path and a torus as in the problem of  ray tracing in computer graphics~\cite{lengyel2012,sup}.

We set  the diameter $\sigma_A$ of $A$  particles  as the unit of length. The simulation box has dimensions  of $Z = 60 $ and $X = Y = 42 $ with membrane thickness $L = 20 $ (Fig.~\ref{fig:method}(a)). Periodic boundary conditions are imposed along all directions such that particles can fly from chamber $1$ to $2$ through the two membranes in the box with opposite streaming velocities $u$ and $-u$. We first pack monodisperse cylindrical pores in parallel on a square lattice with diameters ranging from $d=1.05$ to $3.5$.  The number of pores on the membrane are adjusted to keep the total pore area  $S$ (or porosity) constant. The diameter of the smaller particle $B$ spans over a wide range from $\sigma_B=0.95$ down to $0.005$. We use $\mu_1 = -5$  and $\mu_2 = -5.5, -\infty({\rm particle~ sink})$, which maintain a number density $\rho_1 \lesssim 0.007$ and $\rho_2 \lesssim 0.004$ for each of the two components. The systems under study are thus gaseous for which ideal gas law about pressure holds,   i.e. $p=\rho k_BT$. 
Estimation to the mean free path  $\lambda = \frac{k_B T}{\sqrt{2} \pi \sigma^2 p}$ and Knudsen number ${\rm K_n}   = \lambda / d$ reveals that our systems are within or close to the free molecular flow regime. % (${\rm Kn} > 10^{1}$).

\begin{figure}%[H]%[!b]
\includegraphics[width=0.95\columnwidth]{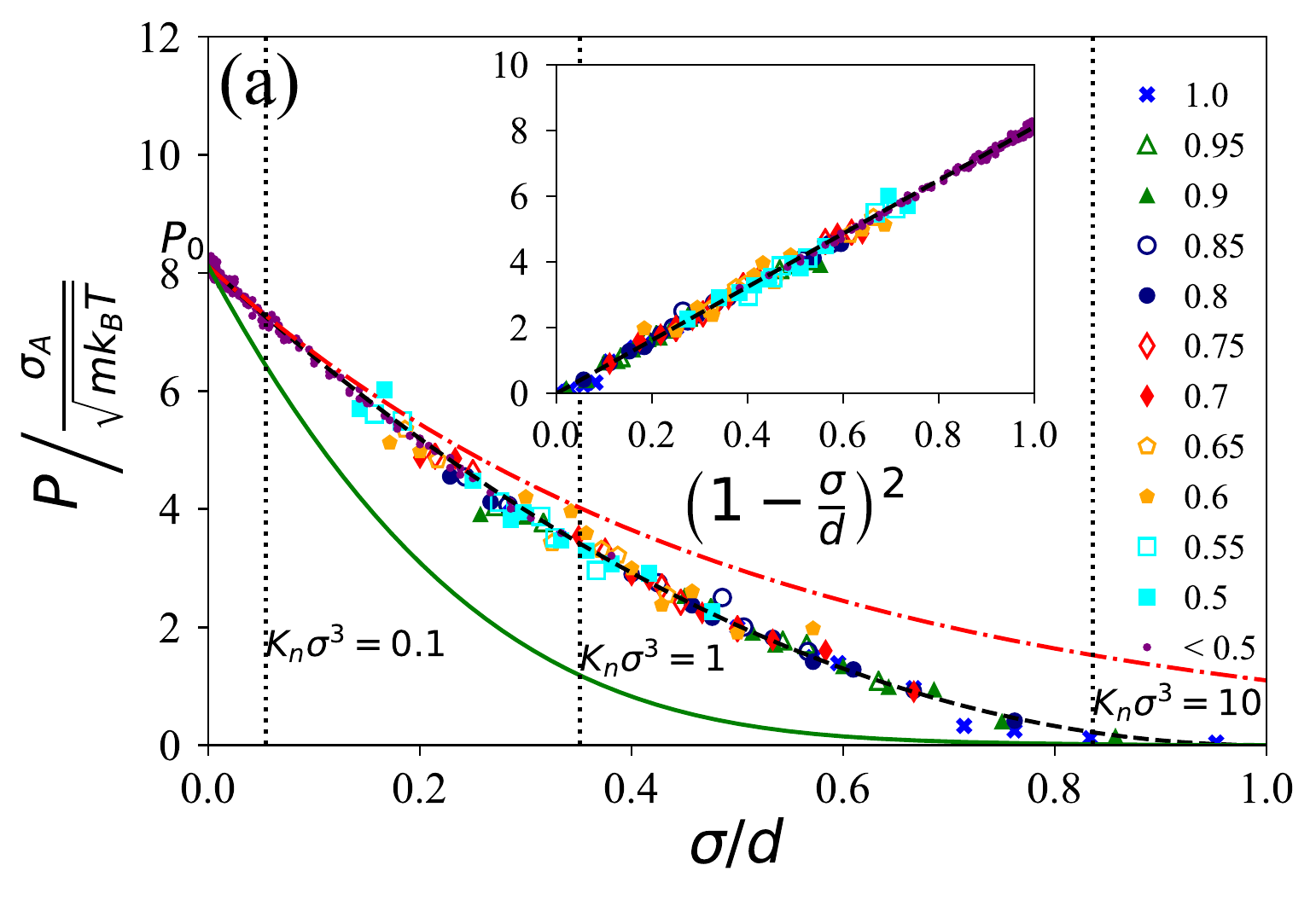}
\includegraphics[width=0.95\columnwidth]{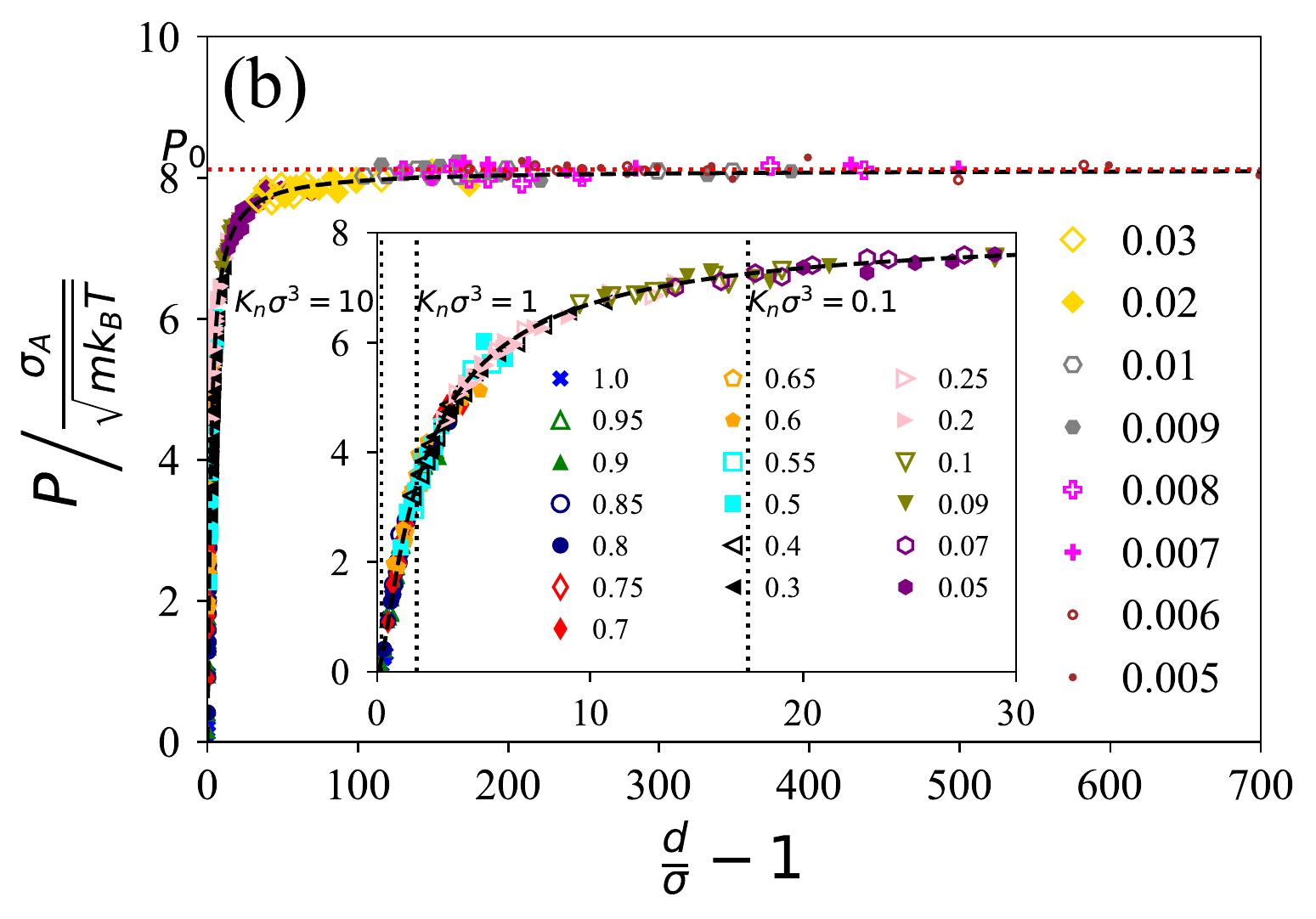}
\caption{Permeability of hard spheres of size $\sigma$ (indicated by different symbols) through pores of diameter $d=1.05-3.5$, following the hindered diffusion law  $P = P_0 (1- \frac{\sigma}{d})^2$ (black dashed). (a)  Exponential law $P = P_0 \exp(-2\sigma/d)$ is approximately obeyed for small $\sigma/d$ (red dotted-dashed).    Renkin's law does not apply here (green solid). (b) $P$ shows   apparent linearity for small $\frac{d}{\sigma} - 1$ but saturates at $P_0$ as $\sigma/d \to 0$. Vertical dotted lines mark the boundaries at ${\rm K_n}\sigma^3=0.1, 1.0, 10$.}
\label{fig:perm}
\end{figure}
The particle flux $J$ (the number of particles going through a membrane per unit pore area per unit time) is calculated from the difference between the net   insertion/removal   numbers in  the two chambers~\cite{heffelfinger1998massively} or from the number of particle crossing events in each pore. Instead of using the transport diffusivity $D$ defined by the Fick's law $J = -D \frac{\Delta \rho}{L}$, we report the transport {\em permeability} $P$ defined by   $J = -P \frac{\Delta p}{L} = -P \nabla p$ and at the low densities of this work  $D \approx P k_BT$.
We find that,  for all the particle sizes $\sigma$ and pore sizes $d$ under study,  $P(d, \sigma)$ collapse onto a universal curve when plotted as a function of the ratio $\sigma/d$ (Fig.~\ref{fig:perm}).  It can be  numerically verified  that  the hindered diffusion law~\cite{beck1970} 
\begin{align}
\label{eq:p}
P(d, \sigma) = P_0 \left( 1- \frac{\sigma}{d} \right)^2
\end{align}
is obeyed with a fitting coefficient $P_0\approx 8$.  The geometric factor  $ \left( 1- \frac{\sigma}{d} \right)^2$ simply originates from the accessible area $\pi(d-\sigma)^2/4$ for a particle to enter the circular opening. Usually, hindered diffusion describes the transport of large particles in liquid solvent~\cite{deen1987hindered}, and due to hydrodynamic effect, an extra factor needs to be included as in the    Renkin's law  $P = P_0 (1- \frac{\sigma}{d})^2   \left (1- 2.104 \frac{\sigma}{d} + 2.09 \frac{\sigma^3}{d^3} - 0.95 \frac{\sigma^5}{d^5} \right )$~\cite{renkin1954}. For solvent-free gaseous particles, the geometric factor alone can determine the permeability (Fig.~\ref{fig:perm}(a)).
When viewed as a function of $d - \sigma$ or $\frac{d}{\sigma} -1$, $P$ shows  apparent linearity when $\sigma$ is comparable with $d$.  But such linearity is just the numerical consequence of the  hindered diffusion law,  i.e. $(1-\sigma/d)^2 = \left[ 1 -\frac{1}{1+(d/\sigma - 1)}\right]^2 = \frac{(d/\sigma - 1)^2}{\left[ 1+(d/\sigma - 1) \right]^2}  $.   Around $\frac{d}{\sigma} -1 \to 0$,  the trend of $P$ is actually curved as $\sim (d/\sigma - 1)^2 $. In the limit of $\sigma \ll d$, instead of having a Knudsen-like diffusion with $P \sim  d$, we observe that $P$ saturates at a constant $P_0$ (Fig.~\ref{fig:perm}(b)).

The front factor $P_0$ in our result can be understood by considering the net flux of point particles through the circular opening at each pore. On average, half of the particles move from chamber 1 to 2 with a streaming velocity $u$ and it takes $\tau = L/u$ for all the $N$ particles inside a pore to be emitted to chamber 2.  Meanwhile, there must be another $N$ particles entering that pore from chamber 1 at steady state, which are originally in the ``waiting'' zone of thickness $L$ next to the pore opening (red shaded area in Fig.~\ref{fig:method}(a)). We can choose the waiting zone to be a cylinder of volume $V$ with a height $L$ and  a radius $\infty$.    The probability that a particle at a given point ${\bf r}$ in this region to enter the pore is determined by the solid angle $\Omega({\bf r})$ made from ${\bf r}$  to the   circular pore opening of an area $s = \pi d^2/4$. If we assume $\rho_1 = \rho$ and $\rho_2 = 0$ for simplicity, then  $N = \rho \int\limits_V {\rm d} {\bf r}  \frac{\Omega({\bf r})}{4\pi} =   \frac{\rho}{4\pi} I(d, L)$, where the integral $I(d, L) =  \int\limits_{0}^\infty r {\rm d}r  \int\limits_{0}^{2\pi} {\rm d}  \theta  \int\limits_{0}^L {\rm d}z \Omega(r,\theta,z) $ can be evaluated numerically for given $d$ and $L$~\cite{paxton1959}. The permeability or diffusivity for point particles can thus be estimated as
$P_0 k_B T=  D_0 = \frac{N}{s\tau} \left/ \frac{\Delta \rho}{L} \right. = \frac{I u}{\pi^2 d^2}$. For $L = 20$, we find $\frac{I }{\pi^2 d^2} = 10.0$ and if we make the approximation that $u \approx \langle |v_z| \rangle = \sqrt{\frac{2k_B T}{\pi m}} \approx 0.798$, $P_0 \approx 7.98$ which is close to the  result of numerical fitting in Eq.~(\ref{eq:p}) .

\begin{figure}%[H]%[!b]
\includegraphics[width=0.95\columnwidth]{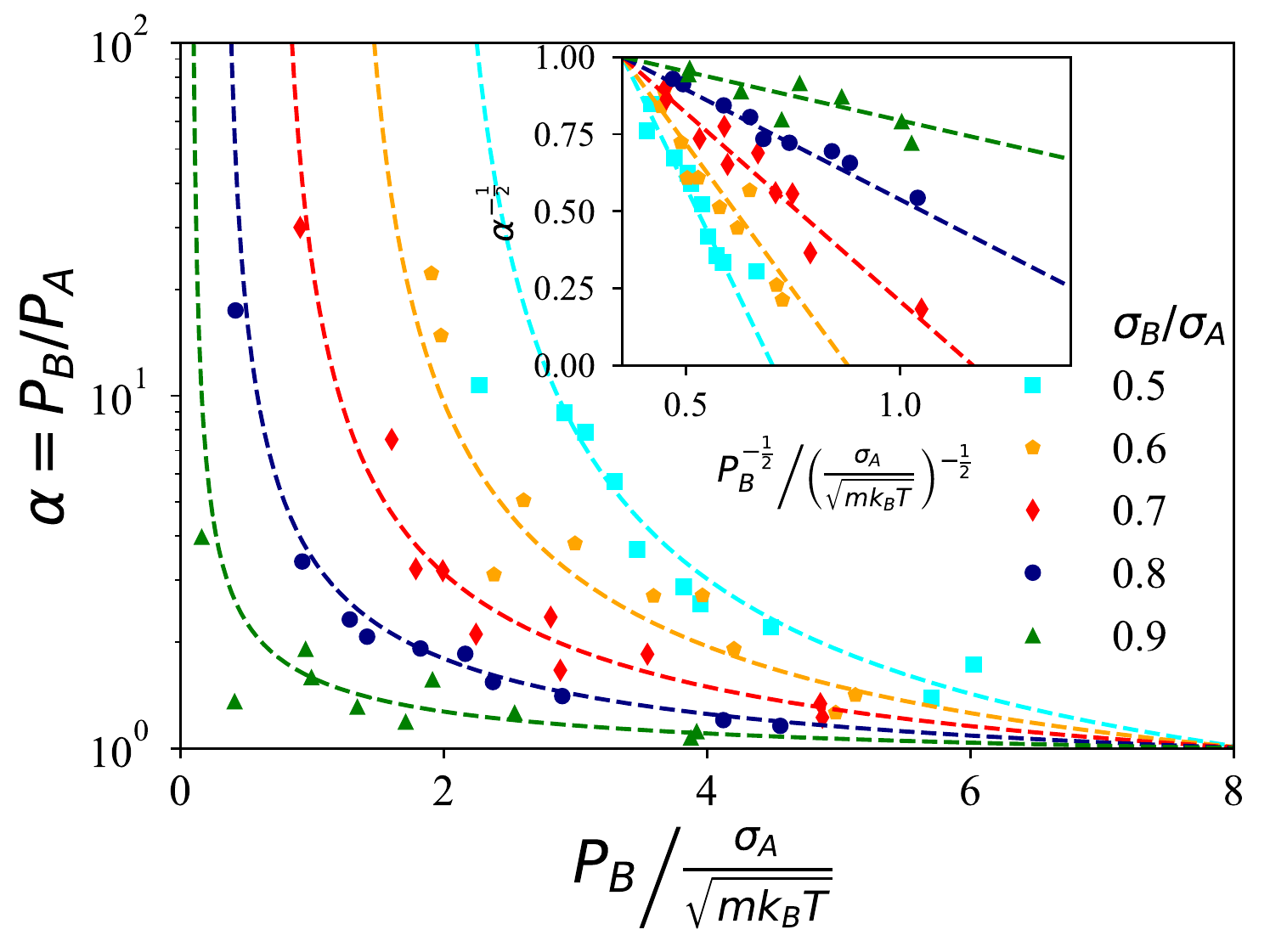}
\caption{Selectivity $\alpha$ versus permeability $P_B$ for binary hard spheres of various size ratios $\sigma_B/\sigma_A$. Inset shows the linear relationship between $\alpha^{-1/2}$ and  $P_B^{-1/2}$.}
\label{fig:alphaP}
\end{figure}

Having confirmed the hindered diffusion mechanism, we can predict the separation curves for binary gaseous mixtures through cylindrical pores as
 \begin{align}
 \label{eq:alpha}
 \alpha = \frac{1}{\left[   \frac{\sigma_A}{\sigma_B}  - \sqrt{P_0} \left( \frac{\sigma_A}{\sigma_B}  - 1\right)    P_B^{-1/2}    \right ] ^2}
 \end{align}
where $\alpha \equiv \frac{P_B}{P_A}>1$ is the (ideal) {\em selectivity} of the more permeable gas $B$ (smaller) with respect to the less permeable one $A$ (Fig.~\ref{fig:alphaP}). In industry, majority of the gas separation membranes are rubbery or glassy polymers, in which gases transport with an activated diffusion mechanism $D \sim \exp( -  \frac{a \sigma^2}{k_BT})$~\cite{ghosal1994,gusev1993}. As a result, the ``upper bound'' of selectivity versus permeability empirically satisfies the linear relationship $\log \alpha = - \lambda_{\rm AB} \log P_B + \kappa$, whose  slope was shown to be $\lambda_{\rm AB} = \lambda_{\rm F} \equiv  \left( \frac{\sigma_A}{\sigma_B}\right)^2 - 1$~\cite{robeson1991,freeman1999basis}. Although there is no strict linear relationship on log-log scale from Eq.~(\ref{eq:alpha}),  it can be shown that, for small $\sigma/d$,  $\log \alpha \approx -\lambda_{\rm AB} \log P_B + \lambda_{\rm AB}  \log P_0$ with $\lambda_{\rm AB } = \frac{\sigma_A}{\sigma_B} - 1$  because    $P\approx  P_0\exp(-2\sigma/d)$. The true linear relationship over the entire regime of $\sigma/d$ is between $\alpha^{-1/2} $ and $P_B^{-1/2}$, i.e.
 \begin{align}
\alpha^{-1/2} = \frac{\sigma_A}{\sigma_B}  - \sqrt{P_0} \left( \frac{\sigma_A}{\sigma_B}  - 1\right)    P_B^{-1/2}.
 \end{align}

\begin{figure}%[H]%[!b]
\includegraphics[width=0.98\columnwidth]{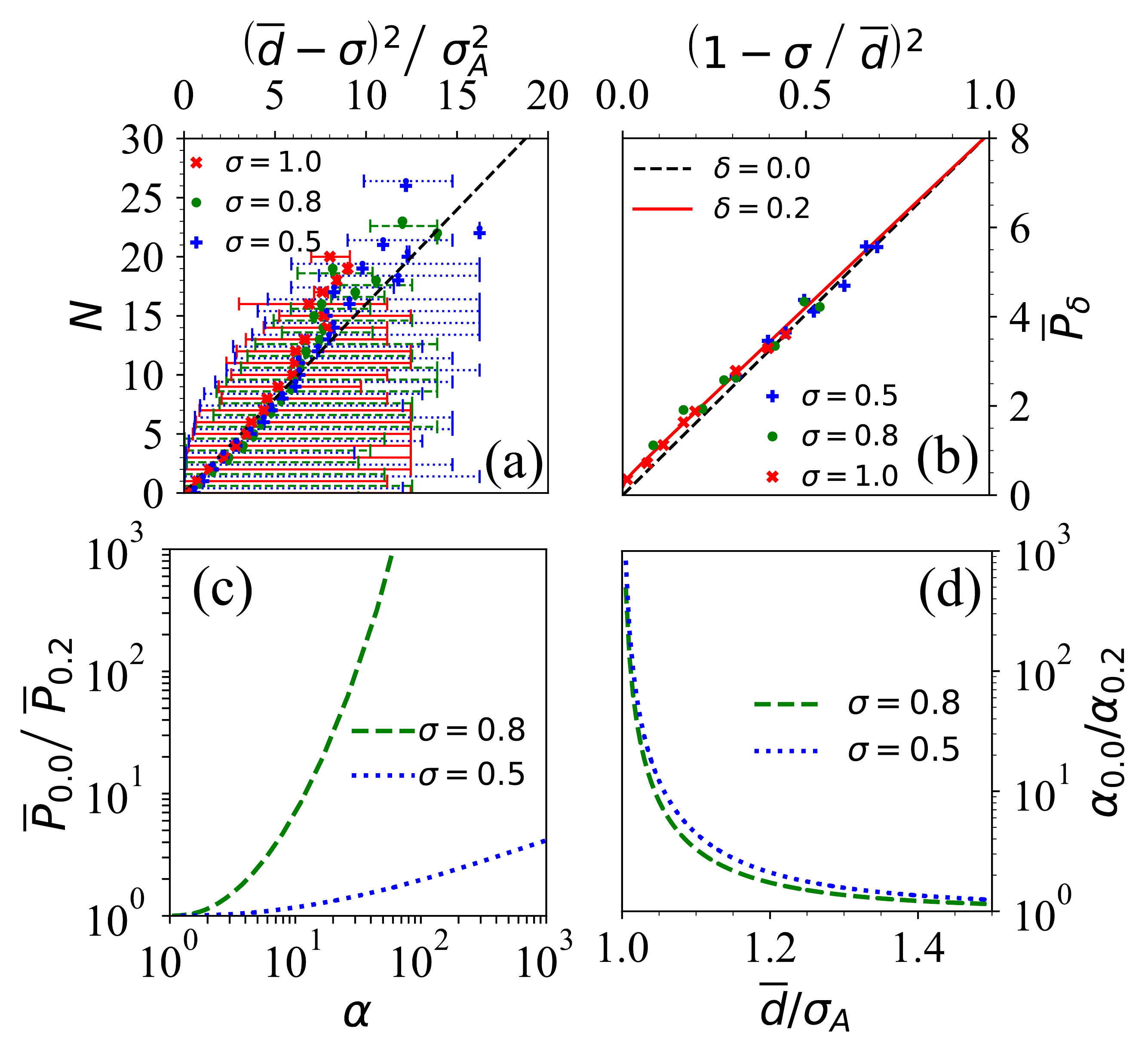}
\caption{(a) Number  $N$ of particles (of size $\sigma_A =1.0$ and  $\sigma_B = 0.8, 0.5$) going through pores of different diameter $d$ during simulations  using polydisperse membranes.  Horizontal line-bars enclose the range of $d$'s for a given $N$ and $\bar{d}$ is their  average.  $N$ is proportional to $(\bar{d} - \sigma)^2$ (black dashed line which is a guide to the eye). (b) The average permeability $\bar{P}_{\delta}$ of the three types of particles going through cylinderical pores with polydispersity $\delta$. Solid line is calculated using the ansatz that the contribution from a pore of size $d$ should be weighted by $d^2$. (c) Calculated permeability enhancement as a function of selectivity and (d) calculated selectivity enhancement as  a function of the mean pore size $\bar{d}$ of the polydisperse membrane, to separate binary mixtures of $\sigma_A = 1$ and $\sigma_B = 0.8,0.5$.}
\label{fig:poly}
\end{figure}

Pore sizes in real membranes are inevitably  polydisperse. We consider variation of $d$ subject to the   
 log-normal distribution $f(d; \bar{b}, \delta)$ with mean $\bar{d}$ and standard deviation $\delta  \bar{d}$ ($\delta$ is called {\em polydispersity})
 \begin{align*}
f(d; \bar{d}, \delta) = \frac{1}{d\sqrt{2\pi\ln(1 + \delta^2)}}e^{-\frac{ \left[ \ln \left(d\sqrt{1 + \delta^2}\left/\overline{d}\right.\right) \right]^2}{2\ln(1 + \delta^2)}}.
 \end{align*}
 At a pressure gradient  $\nabla p$, the total number $N$ of particles going through all pores during time $\tau$ should equal to the sum of number  $N_i$ of particles
 going through each pore $i$ with an area $s_i$, i.e.
 $N = \overline{P}  \nabla p S \tau =  \sum\limits_{i} P_i \nabla p   s_i \tau   =  \sum\limits_{i} N_i $.  If we assume the the permeability for each pore of size $d_i$ is $P_i = P_0 (1-\sigma/d_i)^2$, then $N_i  \propto  P_is_i =  (1-\sigma/d_i)^2 d_i^2 = (d_i - \sigma)^2$. Simulations using polydisperse membranes with $\delta = 0.2$ and $\bar{d} = 1.05$--$3.5$ show that  $N_i $ is indeed proportional to $(d_i - \sigma)^2$ (Fig.~\ref{fig:poly}(a)).
 %\vspace{-0.4 in}
 
 The mean permeability   $ \overline{P} =    \sum\limits_{i} P_i \frac{ s_i}{S}  =   \sum\limits_{i} P_i \frac{ d_i^2}{\sum\limits_{i} d_i^2}$,  for a continuous distribution  $f(d; \bar{d}, \delta)$   of $d$,    can  thus be calculated as
  \begin{align}
  \label{eq:poly}
\overline{P}_{\delta} (\bar{d}, \sigma)  =  \frac{\int_{\sigma}^{\infty}   f(d;\bar{d},\delta) P(d, \sigma)d^2  {\rm d} d }{\int_{\sigma}^{\infty}   f(d;\bar{d},\delta)  d^2  {\rm d} d },
 \end{align}
after knowing the monodisperse permeability $P(d, \sigma)$ as a function of $d$ and $\sigma$ (Fig.~\ref{fig:poly}(b)). Here, the weight  in the average calculation   is $d^2$, which  differs from $d^4$   in  viscous   (Hagen-Poiseuille)  flows~\cite{mehta2005permeability}.  Eq.~(\ref{eq:poly}) can be used to predict permeability enhancement $\overline{P}_{0.0}/\overline{P}_{\delta}$ at a given selectivity   or selectivity enhancement $\alpha_{0.0}/\alpha_{\delta}$ at a given mean pore size $\bar{d}$, when we narrow down the pore size distribution from $\delta>0$ to  $0.0$ (monodisperse) as shown in Fig.~\ref{fig:poly}(c)-(d).

The interests in transport and separation   are not limited to free molecular flows (${\rm K_n} > 10^{1}$),  but have  also been directed to  continuum   (${\rm K_n} < 10^{-3}$), slip   ($10^{-3} < {\rm K_n} < 10^{-1}$) and transition flows ($10^{-1} < {\rm K_n} < 10^{1}$)~\cite{welty2015}.  Decades ago, it was thought that molecular simulations are  computationally too expensive to address these regimes when  the number of particles is large.  Approximate theoretical models or computational methods were thus  developed to tackle these problems, such as the dusty-gas model (DGM)~\cite{mason1967},  the direct simulation Monte Carlo (DSMC)~\cite{bird1963,alexander1997}  and  the finite element method~\cite{roy2003}. With the ever-increasing computer power, it now becomes more and more promising to investigate dense flows on particle level~\cite{plimpton1995,sheng2020}. It would be interesting to extend the current methodolgy to low-${\rm Kn}$ porous systems.\
Caution should be taken  when applying the model used in this work to molecular systems, because the intrinsic roughness of atomic packings, which can induce back reflection and dissipation,  are not captured by smooth cylinderical walls~\cite{malek2003,arya2003knudsen}.

\begin{acknowledgments}
We thank M. Ronen Plesser for helpful discussions on solid angle calculation. We acknowledge  support from Duke Kunshan startup funding. This work
also benefited from  resources made available at the  Duke Compute Cluster (DCC) and the Kunshan Supercomputing Center
(KSSC).
\end{acknowledgments}

%\bibliography{cylinderpore}
%merlin.mbs apsrev4-1.bst 2010-07-25 4.21a (PWD, AO, DPC) hacked
%Control: key (0)
%Control: author (8) initials jnrlst
%Control: editor formatted (1) identically to author
%Control: production of article title (-1) disabled
%Control: page (0) single
%Control: year (1) truncated
%Control: production of eprint (0) enabled
%

\end{document}